\newif\iffigs\figstrue

\documentstyle[12pt]{article}

\iffigs
  \input epsf
\else
\fi

\batchmode
  \newfont{\footscrfont}{rsfs10}
  \newfont{\footbbbfont}{msbm10}
\errorstopmode

\newif\ifscrf\scrftrue
\ifx\footscrfont\nullfont
  \scrffalse
\fi

\newif\ifamsf\amsftrue
\ifx\footbbbfont\nullfont
  \amsffalse
\fi

\def\ppnumber{CLNS-94/1299}
\def\ppdate{September 1994}
\def\pplogo{\vbox{\kern-\headheight\kern -1pt
\halign{##&##\hfil\cr&{
\ppnumber}\cr\rule{0pt}{2.5ex}&\ppdate\cr}
}}

\makeatletter
\date{}
\def\dedicatory#1{\def\@date{\normalsize\it#1}}
\def\subjclass#1{\def\@thefnmark{}\@footnotetext{1991
    {\it Mathematics Subject Classification.} #1}}
\def\keywords#1{\def\@thefnmark{}\@footnotetext{
    {\it Key words and phrases.} #1}}

\def\ps@firstpage{\ps@empty \def\@oddhead{\hss\pplogo}%
  \let\@evenhead\@oddhead 
}
\def\maketitle{\par
 \begingroup
 \def\thefootnote{\fnsymbol{footnote}}
 \def\@makefnmark{\hbox
 to 0pt{$^{\@thefnmark}$\hss}}
 \if@twocolumn
 \twocolumn[\@maketitle]
 \else \newpage
 \global\@topnum\z@ \@maketitle \fi\thispagestyle{firstpage}\@thanks
 \endgroup
 \setcounter{footnote}{0}
 \let\maketitle\relax
 \let\@maketitle\relax
 \gdef\@thanks{}\gdef\@author{}\gdef\@title{}\let\thanks\relax}

\def\abstract{\if@twocolumn
\section*{Abstract}
\else \small
\begin{center}
{\bf ABSTRACT}
\end{center}
\quotation
\fi}

\newif\iffn\fnfalse

\@ifundefined{reset@font}{\let\reset@font\empty}{} 
\long\def\@footnotetext#1{\insert\footins{\reset@font\footnotesize
    \interlinepenalty\interfootnotelinepenalty
    \splittopskip\footnotesep
    \splitmaxdepth \dp\strutbox \floatingpenalty \@MM
    \hsize\columnwidth \@parboxrestore
   \edef\@currentlabel{\csname p@footnote\endcsname\@thefnmark}\@makefntext
    {\rule{\z@}{\footnotesep}\ignorespaces
      \fntrue#1\fnfalse\strut}}}
\makeatother

\ifamsf
  \newfont{\bigbbbfont}{msbm10 scaled\magstep2}
  \newfont{\bbbfont}{msbm10 scaled\magstep1}  
  \newfont{\smallbbbfont}{msbm8}
  \newfont{\tinybbbfont}{msbm6}
  \newfont{\smallfootbbbfont}{msbm7}
  \newfont{\tinyfootbbbfont}{msbm5}
\fi

\ifscrf
  \newfont{\scrfont}{rsfs10 scaled\magstep1}  
  \newfont{\smallscrfont}{rsfs7}
  \newfont{\tinyscrfont}{rsfs7}
  \newfont{\smallfootscrfont}{rsfs7}
  \newfont{\tinyfootscrfont}{rsfs7}
\fi

\ifamsf
  \newcommand{\Bbb}[1]{\iffn
      \mathchoice{\mbox{\footbbbfont #1}}{\mbox{\footbbbfont #1}}
      {\mbox{\smallfootbbbfont #1}}{\mbox{\tinyfootbbbfont #1}}\else
      \mathchoice{\mbox{\bbbfont #1}}{\mbox{\bbbfont #1}}
      {\mbox{\smallbbbfont #1}}{\mbox{\tinybbbfont #1}}\fi}
\else
  \def\bigbbbfont{\bf}
  \def\Bbb{\bf}
\fi

\ifscrf
  \newcommand{\Scr}[1]{\iffn
    \mathchoice{\mbox{\footscrfont #1}}{\mbox{\footscrfont #1}}
    {\mbox{\smallfootscrfont #1}}{\mbox{\tinyfootscrfont #1}}\else
    \mathchoice{\mbox{\scrfont #1}}{\mbox{\scrfont #1}}
    {\mbox{\smallscrfont #1}}{\mbox{\tinyscrfont #1}}\fi}
\else
  \def\Scr{\cal}
\fi

\def\C{{\Bbb C}}

\def\O{{\cal O}}
\def\P{{\Bbb P}}
\def\Q{{\Bbb Q}}
\def\R{{\Bbb R}}
\def\Z{{\Bbb Z}}

\def\opeq#1{\advance\lineskip#1 \advance\baselineskip#1
	\advance\lineskiplimit#1}
\def\eqalignsq#1{\null\,\vcenter{\opeq{2.5\jot}\mathsurround=0pt
	\everycr={}\tabskip=0pt\offinterlineskip
	\halign{\strut\hfil$\displaystyle{##}$&$\displaystyle{{}##}$\hfil
	\crcr#1\crcr}}\,\null}
\def\eqalign#1{\null\,\vcenter{\opeq{2.5\jot}\mathsurround=0pt
	\everycr={}\tabskip=0pt
	\halign{\strut\hfil$\displaystyle{##}$&$\displaystyle{{}##}$\hfil
	\crcr#1\crcr}}\,\null}

\def\sm{$\sigma$-model}

\def\CY{Calabi-Yau}
\def\LG{Landau-Ginzburg}

\def\cM{{\Scr M}}
\def\cA{{\Scr A}}
\def\cB{{\Scr B}}

\def\cMc{{\hfuzz=100cm\hbox to 0pt{$\;\overline{\phantom{X}}$}\cM}}

\def\ff#1#2{{\textstyle\frac{#1}{#2}}}

\def\Vbig{V_{\Delta}}
\def\Vlit{V_\delta}

\begin{document}
\setcounter{page}0
\title{\LARGE On the Geometric Interpretation of \\$N$ = 2 Superconformal
Theories\\[10mm]
}
\author{
Paul S. Aspinwall and Brian R. Greene\\[0.7cm]
\normalsize F.R.~Newman Lab.~of Nuclear Studies,\\
\normalsize Cornell University,\\
\normalsize Ithaca, NY 14853\\[10mm]
}

{\hfuzz=10cm\maketitle}

\def\Large{\large}
\def\LARGE{\large\bf}

\vskip 1.5cm

\begin{abstract}

We clarify certain important issues relevant for the geometric
interpretation of a large class of $N = 2$ superconformal theories. By
fully exploiting the phase structure of these theories (discovered in
earlier works) we are able to clearly identify their geometric
content. One application is to present a simple and
natural resolution
to the question of what constitutes the mirror of a rigid Calabi-Yau
manifold. We also discuss some other models with unusual phase
diagrams that highlight some subtle features regarding the geometric
content of conformal theories.

\end{abstract}

\vfil\break

\section{Introduction and Summary}		\label{s:intro}

One of the most intriguing problems in string theory is to understand
how space-time emerges naturally. Since the vacuum configuration for a
critical string is given by a conformal field theory a question which
arises in this context is the following. Given a conformal field
theory, can one construct some corresponding geometrical
interpretation? In this paper we will discuss this question for
particularly troublesome conformal field theories.
It is worthwhile to emphasize at the outset that in general when
a conformal theory does have a geometrical interpretation it may not
be unique. A perusal of even simple systems such as conformal theories
with central charge $c = 1$ makes this clear. For instance, in this
moduli space it is known that a
string on the group manifold  $SU(2)$ is equivalent
to a string on a circle of radius $\sqrt{\alpha^\prime}$. Both
target spaces have an equal right to be declared {\em the\/} geometrical
interpretation of the conformal field theory. Similarly
a circle of radius $R$ is equivalent to a circle of radius
$\alpha^\prime/R$.
Mirror symmetry, in which strings propagating on distinct Calabi-Yau spaces
give identical physical models, is another substantial arena in which
geometrical interpretations are  not unique.
These  ambiguities are a reflection of the rich structure
of  quantum geometry; they arise because of the extended nature of the
string.

When there are multiple geometric interpretations of a given model,
there is no reason why one should be forced to choose between the
possibilities.
Rather, one can exploit the geometric ambiguity as some interesting
physical questions
are more easily answered from one interpretation rather than another.

In this paper we shall focus our investigation into the geometric content
of certain  of $N = 2$ conformal theories using the framework
established in \cite{W:phase,AGM:I,AGM:II}.  This approach has the virtue
of giving us a physical and mathematical understanding of {\it global\/}
properties of the moduli space of these theories as well as of the theories
themselves. It also gives us the proper arena for
understanding the global implications of mirror symmetry.
We will apply this approach to study some theories whose geometrical
content has been quite puzzling.
For some of these theories, previous papers have proposed
possible geometrical interpretations \cite{Drk:Z,Schg:gen,Set:sup}.
We will see that when phrased in the language of \cite{W:phase,AGM:I,AGM:II},
the previous puzzles are seen to disappear and the geometric status
of these theories becomes apparent.
Following our remarks above, there need not be one unique interpretation
of a given model; however, we do feel that the approach provided here
is especially enlightening and economical. We will also see that the
less natural constructions of \cite{Drk:Z,Schg:gen,Set:sup} can give
misleading results for properties of the corresponding physical model.

We now recall some important background material which will naturally
lead us to a summary of the problems we address and the solutions we offer.

Our understanding of the geometric content of $N = 2, c = 3d$
superconformal theories has undergone impressive growth and revision
over the last few years.  The initial picture which emerged from
numerous studies is schematically given in figure \ref{fig:1}a. We have an
abstract $N = 2, c = 3d$ conformal field theory moduli space that is
geometrically interpretable in terms of complex structure and K\"ahler
structure deformations of an associated Calabi-Yau manifold of $d$
complex dimensions and a fixed
topological type. The space of K\"ahler forms naturally exists as a
bounded domain (the complexification of the ``K\"ahler cone'') which
we denote as a cube. The moduli space of complex structures does not
have this form and is more usually compactified to form a compact space.
Observables in each of the conformal theories in the
moduli space are related to geometrical constructs on the
corresponding Calabi-Yau space, the latter being taken as the target
space of a nonlinear sigma model.

\iffigs
\begin{figure}
  \centerline{\epsfxsize=9cm\epsfbox{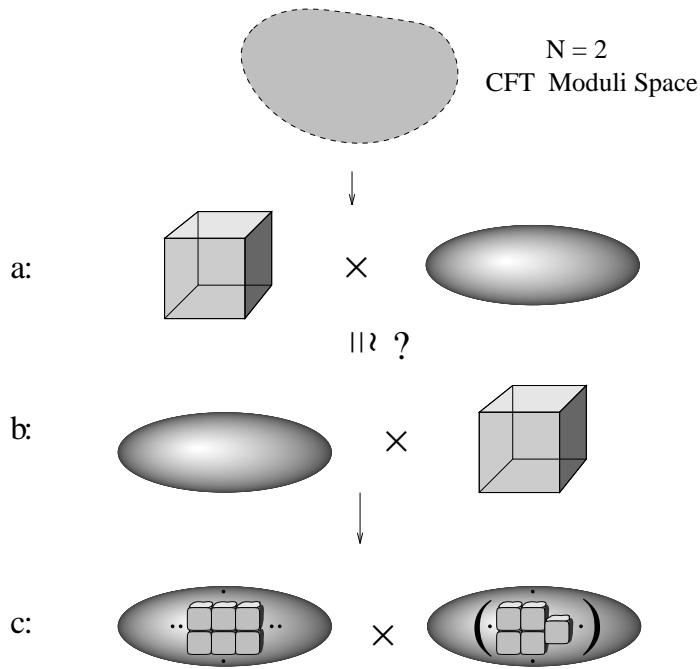}}
  \caption{Models of the moduli space.}
  \label{fig:1}
\end{figure}
\fi

This picture was extended to that given in figure \ref{fig:1}b after the
discovery of mirror symmetry.  Two Calabi-Yau spaces $X$ and $Y$
constitute a mirror pair if they yield isomorphic conformal theories
when taken as the target space for a two-dimensional supersymmetric
nonlinear \sm, with the explicit isomorphism being a change in
sign of the left moving $U(1)$ charges of all fields. Geometrically
this implies that the Hodge numbers $h^{1,1}(X)$ and $h^{d-1,1}(X)$ are
related to those of $Y$ by $h^{1,1}(X) = h^{d-1,1}(Y)$ and
$h^{d-1,1}(X) = h^{1,1}(Y)$.  Since the cohomology groups $H^{1,1}$
and $H^{d-1,1}$ correspond to K\"ahler and complex structure
deformations, respectively, we see that the underlying conformal field
theory moduli space has the two geometrical interpretations given in
the figure. This immediately led to a problem since, as mentioned
above, the geometric form of the moduli spaces of K\"ahler forms and
complex structures appeared to be quite different.

This was resolved by the works of
\cite{W:phase,AGM:I,AGM:II} to that shown in figure \ref{fig:1}c. Here
we see that the appropriate
interpretation of the conformal field theory moduli space has required
that the K\"ahler moduli space of $X$ be replaced by its ``enlarged
K\"ahler moduli space'' (and similarly for $Y$).  The latter contains
numerous regions in addition to the K\"ahler cone of the topological
manifold $X$. For instance, it typically contains regions
corresponding to the K\"ahler cones of Calabi-Yau spaces related to
$X$ by the birational operation of flopping a rational curve, regions
corresponding to the moduli space of singular blow-downs of $X$ and
its birational partners, and regions interpretable in terms of the
parameter space of (gauged or ungauged) Landau-Ginzburg models fibered
over various compact spaces. The complex structure moduli space can
also be equipped with a phase structure
\cite{AGM:sd} --- as must happen to
preserve mirror symmetry. We note that from the \sm\ point of view
the phase regions in the complex structure moduli space have a less
pronounced physical interpretation. This is because in analyzing the \sm\
we use perturbation theory in K\"ahler modes (which fix
the size of the Calabi-Yau) and hence this approximation method is not mirror
symmetric. However, the phase structure in the complex structure
moduli space of $X$
{\it is\/} the phase structure in the enlarged K\"ahler moduli space
of $Y$ and
it is the latter interpretation where this phase structure is most manifest.
For the purposes of this paper we may ignore the phase structure in
the complex structure part of the moduli space and
for this reason we have put parentheses around this in \ref{fig:1}c.

The results of the present paper all stem directly from a careful
study of the phase diagrams of figure \ref{fig:1}c. We shall review the
quantitative construction of these phase spaces in section \ref{s:ph}; for now
we will content ourselves with the schematic description given and
summarize our results with a similar level of informality.

There are numerous ways of constructing $N = 2$ superconformal
theories with $c = 3d$. Some constructions, such as the Calabi-Yau
\sm s described above, are manifestly geometric in
character. Other constructions do not begin with a geometrical target
space and hence their geometrical content, if any, can only be
assessed after more detailed study. More generally and pragmatically,
given an abstract conformal field theory in some presentation, how do
we determine if it has a geometrical interpretation?  We will not seek
to answer this question in generality, but rather will focus attention
on those theories for which we can construct the phase diagram
illustrated in figure \ref{fig:1}c. For theories of this sort, as we shall
review, toric geometry supplies us with a geometric description
of each theory. We hasten to emphasize, though, that Calabi-Yau \sm s
are but one kind of corresponding geometry. We will see, for instance,
that Landau-Ginzburg orbifolds  can be associated with noncompact,
generally singular, configuration spaces.
%
%
%

{}From our brief discussion here and also from
\cite{W:phase,AGM:I,AGM:II}
 one might think that any theory with a phase diagram such as that
in figure \ref{fig:1}c, has regions interpretable in terms of
Calabi-Yau \sm s. After
all, our
progression from figures \ref{fig:1}a through \ref{fig:1}c has centered around
K\"ahler cones of Calabi-Yau spaces.  This conclusion, as we
shall see in detail in section \ref{s:appl}, is false and comes to bear on a
number of issues, including that of the generality of mirror
symmetry. Namely, there are Calabi-Yau manifolds that are rigid,
i.e. that have trivial $H^{d-1,1}$. The mirror to such a space,
therefore, should have $h^{1,1} = 0$. This is troublesome, though,
because Calabi-Yau spaces are K\"ahler and hence have at least one
nontrivial element in $H^{1,1}$.  With the above discussion, and
explicit calculation in section \ref{s:appl}, the resolution to this puzzle
becomes clear: the enlarged K\"ahler moduli space for the theory
mirror to the one associated with the rigid space $X$ {\it does not
contain a region interpretable in terms of
a Calabi-Yau \sm}.  In fact, the enlarged K\"ahler moduli space, in
contrast to the generic case illustrated in figure \ref{fig:1}c, is
zero-dimensional and consists of a single point. By direct analysis,
we show that the corresponding theory is a Landau-Ginzburg orbifold -
not a Calabi-Yau \sm\ - and hence it is perfectly consistent
for the theory to lack a K\"ahler modulus.  We note at the outset that
possible resolutions to the question of the identity of mirrors to
rigid Calabi-Yau spaces have been previously presented in
\cite{Drk:Z,Schg:gen,Set:sup}. These authors have invoked unexpected
additional structures
such as non-Calabi-Yau spaces of dimension greater than $d$ and
supermanifolds in an attempt to resolve this issue. Contrary to these
works, we see here that absolutely no additional structure is
required. Rather, rigid Calabi-Yau manifolds fit perfectly into the
general framework introduced in \cite{W:phase,AGM:I,AGM:II}.

In addition to applying our analysis to the case of rigid Calabi-Yau
manifolds and their mirrors, we also study two other interesting
phenomena. First, we present an example of a theory with
nonzero dimensional enlarged K\"ahler moduli space that does not
contain a geometric region thus showing that the mere existence of a
would-be K\"ahler form does not guarantee a Calabi-Yau
interpretation. Second, we briefly discuss an example (first pointed
out in \cite{W:phase}) whose enlarged K\"ahler moduli space has a
phase whose target space has the desired dimension but is not of the
Calabi-Yau type.


\section{Phase Diagrams: Supersymmetric Gauge Theory and Toric Geometry}
	\label{s:ph}

The moduli space of $N = 2$ superconformal theories is most naturally
interpretable in terms of a collection of regions within which the theory
assumes a particular phase. Amongst the possibilities are  smooth and singular
geometric Calabi-Yau phases, gauged and ungauged Landau-Ginzburg phases,
as well as orbifolds and hybrids thereof. This is the burden of figure
\ref{fig:1}c.

The existence and quantitative construction of these phase diagrams has been
approached from two distinct vantage points in the works of
\cite{W:phase} and \cite{AGM:II}. In fact, a point which is not as fully
appreciated as it might be is that these two approaches, although phrased
in different languages, are {\it isomorphic}. Different
questions, though, are often more easily answered from one of the two
formalisms and hence it is important to fully understand both
approaches and their
precise relationship.  It is the purpose of the present section to
explain these issues. We note that the material in this section is
implicit in \cite{W:phase} and \cite{AGM:II}; for our purposes we need to
make the relation explicit.

In brief, both \cite{W:phase} and \cite{AGM:II} build constrained $N = 2$
supersymmetric quantum field theories. In the physical approach of
\cite{W:phase}
these constraints are phrased in terms of {\it symplectic} quotients.
In the mathematical approach of  \cite{AGM:II} these constraints are phrased
in terms of {\it holomorphic} quotients. The well-known equivalence
\cite{Kirwan:,Ness:} of
these two approaches then implies that each constructs the same theory and
hence also the same phase diagrams. The proper language for establishing
these statements is that of {\it toric geometry} for which the reader can
find a primer in \cite{AGM:II}. In the following we will try to convey the
main points with a minimum of unnecessary technical detail.

Complex projective space may be considered to be the prototypical
toric variety. One constructs $\P^n$ by taking the $n+1$ homogeneous
coordinates, $x_i$, spanning $\C^{n+1}$, removing the origin $x_i=0$
and modding out by the $\C^*$-action $x_i\to\lambda x_i$,
$\lambda\neq0$. A toric variety is simply a generalization of this
concept with perhaps more than one $\C^*$-action and a possibly more
complicated point set removed prior to the modding out process.

The most natural way of building a $N$=2 \sm\ with a complex
projective target space appears to be in terms of a $U(1)$-gauged
field theory \cite{ADL:CPn}. In this construction, one begins with the
homogeneous coordinates, $x_i$, denoting chiral superfields, each with
the same $U(1)$ charge, $Q_i$, (which we may take to be 1). The
classical vacuum of such a theory may be determined by finding the
minimum of the classical potential energy. Solving the algebraic
equations for the auxiliary D-component of the gauge
multiplet  and including the result in the scalar
potential yields the familiar contribution
\begin{equation}
 ( |x_1|^2 + |x_2|^2 +\ldots+|x_{n+1}|^2 - r )^2,	\label{eq:ve1}
\end{equation}
where $r$, a real number, is the coefficient of the familiar Fayet-Illiopoulos
D-term. We take $r$ to be positive here to avoid na\"\i vely breaking
supersymmetry (see section 3.2 of \cite{W:phase} for a discussion on
negative values of $r$).
Minimizing the energy forces us to require that (\ref{eq:ve1}) should
vanish. This
immediately removes the origin $x_i=0$ from consideration. It also
forces the $x_i$ to lie on the sphere $S^{2n+1}$. We may now divide
out by the $U(1)$ (i.e., $S^1$) action to form
$S^{2n+1}/S^1\cong\P^n$. The process
of dividing by $\C^*$, in the usual formulation of
$\P^n$ may be viewed to having taken place in two
stages. First we fix an $\R_+$ degree of freedom by imposing the
vanishing of (\ref{eq:ve1}) and then we divide out by $S^1$. The
equivalence of these two constructions
then follows from the fact that $\C^*\cong \R_+\times S^1$. Dividing by
the former
is a simple example of a holomorphic quotient; dividing by the latter is
a simple example of a symplectic quotient. We have just seen, therefore,
the essential reason why these two are equivalent. Let us now discuss
how Witten generalized upon this quantum field theory approach of
generating symplectic quotients. We will then discuss their
equivalent holomorphic
quotient description as in  \cite{AGM:II}.

Witten \cite{W:phase} extended the above model to describe not a
complex projective space but the ``canonical'' line bundle of
complex projective space (see, for example, \cite{GH:alg} for the precise
definition of this bundle). Let us reserve $n$ to denote the dimension
of the toric variety in question so now we are looking at a line
bundle over $\P^{n-1}$ and the variable $x_{n+1}$ will now be treated
differently to the others.
This space is then built from $\C^{n+1}$ by removing the point
$x_1=x_2=\ldots=x_n=0$ and modding out by the action
\begin{equation}
\eqalign{
  x_i&\to \lambda x_1,\quad i=1\ldots n,\cr
  x_{n+1}&\to\lambda^{-n}x_{n+1}.\cr}	\label{eq:Cs1}
\end{equation}
To produce this from the gauged \sm\ point of view we put $Q_i=1$ for
$i=1\ldots n$ and $Q_{n+1}=-n$.

The vanishing of the classical potential now implies
\begin{equation}
  |x_1|^2 + |x_2|^2 +\ldots+|x_n|^2 -n|x_{n+1}|^2 = r.	\label{eq:ve2}
\end{equation}
We see that there are classical vacuum solutions for $r$ for
either sign.
If $r>0$, we thus recover the required target space as in the case of
the projective space. If however $r<0$, we find that $x_{n+1}\neq0$
and we have no condition on $x_1,\ldots, x_n$. Let us consider this
space more closely.

Removing the point set
$x_{n+1}=0$ from $\C^{n+1}$ and dividing by the
action (\ref{eq:Cs1}) produces another toric variety. $x_{n+1}$ may be
fixed by choosing a value for $\lambda^n$ leaving the $n$th roots of
unity to act on the space spanned by $x_1,\ldots,x_n$. Thus the toric
variety is $\C^n/\Z_n$. Therefore we see that the geometry of the
target space can change discontinuously as we vary $r$. This theory is
said to have two {\em phases\/} where the relevant toric variety is
either the canonical line bundle of $\P^{n-1}$ or $\C^n/\Z_n$.

The construction of \cite{W:phase} doesn't quite stop here. One may
introduce a $U(1)$-invariant superpotential, $W$, i.e., a $\C^*$-invariant
polynomial over the $x_i$'s. Minimizing the classical potential now
also implies that we are at a critical point of $W$.

Our toric ``ambient'' space will always turn out to be non-compact.
This however will contain compact subspaces which may also be considered as
toric varieties themselves. Clearly $\P^{n-1}$ is a toric subspace of
the canonical line bundle over $\P^{n-1}$. The only compact toric
subspace of $\C^n/\Z_n$ is the point at the origin. Assuming that $W$
is suitably generic, the effect of
including the superpotential term is to force the classical vacuum to
be equal to, or contained in some compact toric subspace of the
ambient space.

In our example, a suitable $W$ is $(x_1^n+x_2^n+\ldots+x_n^n)x_{n+1}$.
In the canonical line bundle over $\P^{n-1}$ case, the critical point
set of $W$ consists of the hypersurface $x_1^n+x_2^n+\ldots+x_n^n=0$
in $\P^{n-1}$. This is a compact \CY\ $(n-2)$-fold. This is thus named, the
{\em \CY\/} phase. In the $\C^n/\Z_n$ case, the origin is the critical
point set of $W$. Thus our classical vacuum is simply one point. The
effective superpotential of this theory however allows for massless
fluctuations around this point given by a Landau-Ginzburg
superpotential $x_1^n+x_2^n+\ldots+x_n^n$. This is thus the {\em\LG\/}
phase. Note that all fluctuations around the vacuum in the \CY\ phase
are massive.

Let us fix some notation.\footnote{This notation is not entirely
consistent with \cite{AGM:II}. For example the $\Delta$ of this paper
is the $\Delta^+$ of \cite{AGM:II}.} We will call the ambient non-compact
toric space $\Vbig$. This contains a maximal compact toric subset
$\Vlit$ (which may be reducible). Within $\Vlit$ we have the classical
vacuum of the quantum field theory which we denote $X$.

A simple generalization of the above construction is to consider a
weighted projective space for $\Vlit$. Clearly this may be achieved by
giving different charges to $x_1,\ldots,x_n$. Following the above
formalism we would again obtain two phases depending on whether $r$
was less than or greater than zero.
When we look at the associated conformal field theory
 it turns out that this does not
capture the full moduli space, i.e., $h^{1,1}>1$ for many of these theories.
It is not hard to generalize the present description
to include at least some of these other degrees of freedom. For each
such independent
direction in the moduli space we are able to access in this formalism,
we introduce a $U(1)$ gauge factor and
a corresponding parameter $r_l$. Thus, the total gauge group is
$G=U(1)^{s}$ where $s$ is the dimension of this subspace of the moduli
space .
The chiral fields
will in general be charged under all of the $U(1)$ factors, and hence we
write $Q_i^{(l)}$ to denote the charge of the $i^{th}$ chiral superfield under
$U(1)_{(l)}$. The superpotential $W$ must now be a $G$-invariant
combination of the chiral superfields.

It turns out that the language of toric geometry is precisely suited for
determining all of the data needed for building such a model. Namely,
in the case of $s = 1$ (or more generally, $s$ is the number of distinct
 toric factors making up the ambient space) it is straightforward
to figure out appropriate charges so that minimization of the scalar potential
yields the desired model. When $s$ is not of this form, the problem requires
a more systematic treatment; this is precisely what the formalism of toric
geometry supplies. Furthermore, for these more general cases, it proves
increasingly difficult to determine the phase diagram of the model by
studying the minimum of the scalar potential for various values of
the $r_1,...,r_{s}$.  The formalism of toric geometry,
as described in  \cite{AGM:II},  supplies us with a far more efficient
means of determining the phase structure, as well. Hence,
 let us  now recast the above formulation directly in
terms of toric geometry.

The homogeneous coordinates (in the sense of \cite{Cox:})
 $x_1,\ldots, x_N$
form a natural representation of the group $(\C^*)^N$. Let us
form a toric variety by removing some point set and dividing the
resultant space by $(\C^*)^{N-n}$. Clearly the space formed, $\Vbig$,
is acted upon non-trivially by $(\C^*)^n$.
Let us introduce $\zeta_j$, $j=1,\ldots,n$, as the natural
representation of this $(\C^*)^n$-action. That is, the $\zeta_j$ provide
coordinates on a dense open subset of $\Vbig$. This follows since
$\Vbig$ may be regarded itself as a compactification of $(\C^*)^n$.
Let us relate these new ``affine'' coordinates to the homogeneous
coordinates by
\begin{equation}
  \zeta_j = \prod_{i=1}^N x_i^{\alpha_{ij}},	\label{eq:aff}
\end{equation}
where $\alpha_{ij}\in\Z$.
We may represent the $N\times n$ matrix, $\alpha_{ij}$, by a collection of
$N$ points, which we denote $\cA$,
living in an $n$-dimensional real space where
$\alpha_{ij}$ is the $j$th coordinate of the $i$th point.
Let us demand that $\cA$ is such that there exists an $n$-dimensional
lattice ${\bf N}$ within this same
space (which we denote ${\bf N}_{\R}={\bf N}\otimes_{\Z}\R$) such that
\begin{equation}
  \cA = {\bf N} \cap (\hbox{Convex hull of $\cA$}\,).
		\label{eq:Acvx}
\end{equation}
The notation
$\alpha_i$ will denote the position vector of the $i$th point of $\cA$
in ${\bf N}$.

Consider now the charges of the homogeneous coordinates under the
$(\C^*)^{N-n}$ by which we modded out. Denote these $Q_i^{(l)}$ where
$i=1,\ldots,N$ and $l=1,\ldots,N-n$. The obvious short exact sequence
\begin{equation}
  1\to(\C^*)^n\to(\C^*)^N\to(\C^*)^{N-n}\to1,
\end{equation}
induces,
\begin{equation}
  \sum_{i=1}^N Q^{(l)}_i\alpha_{ij} = 0,\quad\forall l,j.
	\label{eq:krnl}
\end{equation}

Thus, we see that the charges $Q^{(l)}_i$ are simply the {\it kernel of
the transpose of the matrix whose elements are\/} $\alpha_{ij}$. The
reader should check
that in the simple case, say, of projective space discussed earlier,
that the charge assignment posited can in fact be derived in this
manner.

Now define ${\bf M}$ as the dual lattice to ${\bf N}$. Let us demand
that there is an element $\mu\in{\bf M}$ such that
\begin{equation}
  \langle\mu,\alpha_i\rangle=1,\quad\forall i.	\label{eq:hypln}
\end{equation}
This condition is similar to stating that $\Vbig$ be a space with
vanishing canonical class, $K$, (or zero first Chern class). Actually $\Vbig$
need not be smooth so be need to be more careful about our language.
The correct term from algebraic geometry is that $\Vbig$ is
{\em Gorenstein\/} (see, for example, \cite{Reid:mm}). Applying
(\ref{eq:hypln}) to (\ref{eq:krnl}) tells us that
\begin{equation}
  \sum_{i=1}^N Q^{(l)}_i =0,\quad\forall l.  \label{eq:Q0}
\end{equation}
This appears as an important condition in \cite{W:phase}
ensuring freedom from anomalies in certain chiral currents
which should be present if there is an infrared limit with $N$=2
superconformal invariance.
It is curious to note that (\ref{eq:Q0}) is not sufficient to guarantee
(\ref{eq:hypln}). We may have $\langle\mu,\alpha_i\rangle=k$ for
example, for some integer $k$. $\Vbig$ would then be $\Q$-Gorenstein
which is roughly saying that $kK=0$ but $K$ may be a non-trivial
torsion element. The effect of this in terms of the two dimension
quantum field theory has not been studied.

This point set $\cA$ gives us all the information we require to build
$\Vbig$ except which point set should be removed from $(\C^*)^N$
before performing the quotient. This is performed in toric geometry by
building a fan, $\Delta$. A fan is a collection of tesselating
cones in ${\bf N}_\R$ with apexes at the origin. The
intersection of this fan with the hyperplane containing $\cA$ will be
a set of tesselating polytopes. The convex hull of this set of
polytopes must be the convex hull of $\cA$ and the vertices of the
polytopes must be elements of $\cA$. Thus each cone, $\sigma$, in
$\Delta$ is ``generated'' by a subset of $\cA$. We say
$\alpha_i\in\sigma$ if $\alpha_i$ is one of the generators, i.e.,
$\alpha_i$ lies at a vertex of the intersection of $\sigma$ with the
hyperplane in ${\bf N}_\R$ containing $\cA$.
The point set $F_\Delta$ removed from $\C^N$ prior to quotienting is
then specified by
\begin{equation}
  \bigcap_{\sigma\in\Delta} \Bigl\{ x\in\C^N;
  \prod_{{\alpha_i\in\cA,} \atop {\alpha_i\not\in\sigma}}\!\!x_i=0 \Bigr\},
		\label{eq:Fset}
\end{equation}
where $x$ is the point with coordinates $x_i$.

The fact that different fans may be associated with the
point-set $\cA$ gives rise to the phase structure. We need only
consider the case where all the $\sigma$'s are simplicial based cones,
i.e., we induce a simplicial decomposition of triangulation of $\cA$.
To each such fan (satisfying
in addition a certain ``convexity'' property, see \cite{AGM:II} for
more details) we associate a phase. Other fans consistent with $\cA$
not satisfying these conditions
may always be considered as models on the boundary between two or more
phases. The parameters, $r$, in the linear \sm\ approach give us an
identical fan structure. This is best understood from examining figure
11 of \cite{AGM:II}. The $r$ parameters, in essence, fix the heights
of the points in this figure and hence following the discussion of
section of 3.8 of \cite{AGM:II} their values determine a triangulation
of the point set $\cA$. From a physical point of view we can group
together those values for the $r$ parameters which yield the same
phase for the model. In this way we partition the space of all
possible $r$'s into a phase diagram. This phase diagram is the
``secondary fan'' for the moduli space as discussed in \cite{AGM:II}.

We now have a dictionary between \cite{W:phase} and the toric
approach: {\em Specifying generic values of {\rm ``$r$''} parameters is
equivalent to specifying a triangulation of $\cA$. The non-vanishing
conditions on the fields $x_i$ specified by minimizing the $D$-term
part of the classical potential is equivalent to removing the point
set $F_\Delta$ given by\/} (\ref{eq:Fset}).

Note that requiring $\cA$ to be ``complete'' in the sense of
(\ref{eq:Acvx}) is not necessary in the analysis of \cite{W:phase}. By
imposing this condition we gain access to the largest subspace of the
moduli space we can reach by this toric method.

\def\cBp{\cB\,{}^\prime}
One point in the dictionary between \cite{W:phase} and the toric
approach which we have not spelled out explicitly as yet is how we
determine the
superpotential $W$ from the toric data. This is straightforward as
we now describe. Let us $G$ to denote the group $(\C^*)^{N-n}$.
$W$ is a $G$-invariant polynomial in the chiral
superfields.  From (\ref{eq:krnl}) we see that any monomial of the
form
\begin{equation}
\prod_{i=1}^N x_i^{\langle\alpha_i, v\rangle} \label{eq:prod}
\end{equation}
for a fixed but arbitrary vector $v$ is $G$-invariant.
However, we want all terms in $W$ to not only be $G$-invariant but
also to have nonnegative integral  exponents. Towards this end
we are naturally led to introduce  the  cone $\Upsilon$ in ${\bf M}_\R$,
dual to  $\Sigma$ which is the cone over the convex hull of $\cA$ in
${\bf N}_\R$,
defined by
\begin{equation}
  \Upsilon = \left\{ s\in{\bf M}_\R; \langle s,t\rangle\geq0, \forall
	t\in\Sigma \right\}.
\end{equation}
The integral lattice points in $\Upsilon$, when substituted for the vector
$v$ in (\ref{eq:prod}), will then generate $G$-invariant
monomials with nonnegative exponents. To systematize this,
we now define $\cBp\subset \Upsilon$ by
\begin{equation}
 \cBp = {\bf M} \cap \Upsilon,
\end{equation}
the integral lattice points contained in the dual cone.
Any point in $\cBp$, if substituted for the vector $v$ in (\ref{eq:prod}),
yields a $G$-invariant nonnegative exponent monomial.
Finally, we note that we would like $W$ to be a suitably
``quasihomogeneous'' polynomial
of lowest nontrivial degree in the $x_i$. This will remove any
``irrelevant'' terms in the superpotential \cite{VW:} and may be
achieved as follows.
Let the monomials in this reduced superpotentials be labeled by
elements of $\cB \subset \cBp$.
Following \cite{BB:mir} let us put one last condition on
$\cA$, namely that when we derive the point set $\cB$
exists a vector $\nu\in{\bf N}$ such that
\begin{equation}
  \langle \beta_v,\nu\rangle = 1, \quad\forall\beta_v\in\cB,
\end{equation}
and that the vectors given by the elements of $\cB$ (or a subset of
$\cB$) generate $\Upsilon$.
We also impose the condition on $\cB$ paralleling our discussion for
the point set $\cA$.
Namely, we can say
\begin{equation}
  \cB = {\bf M} \cap (\hbox{Convex hull of $\cB$}\,),
\end{equation}
with
the elements of $\cB$ at the vertices of this convex hull generating
$\Upsilon$.
We denote by $M$ the number of points in $\cB$ so that $v=1,\ldots,M$.
The superpotential $W$ is then constructed according to
\begin{equation}
  W = \sum_{v=1}^M a_v w_v,
\end{equation}
for $a_v\in\C$
with
\begin{equation}
  w_v = \prod_{i=1}^N x_i^{\langle\beta_v,\alpha_i\rangle}.
			\label{eq:mon}
\end{equation}

We may note at this point that mirror symmetry is conjectured to
exchange the sets $\{{\bf M},\mu,M,\cA\,\}\leftrightarrow\{{\bf
N},\nu,N,\cB\}$. This may be regarded as a generalization of the
``monomial-divisor mirror map'' of \cite{AGM:mdmm}.\footnote{This has
also been studied independently by S. Katz and D. Morrison
\cite{KM:mir}.}
The mirror pairs of \cite{GP:orb} (which is established at the conformal
field theory level) are a subset of this general construction
and the examples in sections \ref{ss:Z} and \ref{ss:h1} are in this
subset. Thus statement concerning mirror symmetry with regards to
these examples may be regarded as definitely true. Also note that our
analysis of the phases of the moduli space does not depend on the
mirror map and thus does not depend on this mirror conjecture.

Now let us try to calculate the central charge $3d$ of the conformal
field theory associated to this model. We may apply the same reasoning
as was used in \cite{VW:} to determine this. Firstly we have $N$
chiral superfields each of which contributes $+1$ to $d$. This may be
taken to correspond to the string propagating in $\C^N$. We also have $N-n$
vector superfields which we take to contribute $-1$ to $d$ since each
removes one complex dimension from the target space. Thus, so far we
have $d=n$. However, the string is further confined by the
superpotential $W$ and we expect this to reduce the value of $d$ as we now
show.

Consider now rescaling by an element of $(\C^*)^N$,
\begin{equation}
  x_i\to\lambda^{\omega_i}x_i.
\end{equation}
The monomial $w_v$ then scales to $\lambda^\chi w_v$ where
\begin{equation}
  \chi=\sum_{i=1}^N \langle\beta_v,\alpha_i\rangle\omega_i.
\end{equation}
Consider choosing the weights $\omega_i$ such that
\begin{equation}
  \sum_{i=1}^N \omega_i\alpha_i = \nu. \label{eq:wcond}
\end{equation}
Then all the monomials transform $w_v\to\lambda w_v$ and thus,
declaring $a_v$ to be invariant, we have $W\to\lambda W$. Taking the
inner product of (\ref{eq:wcond}) with $\mu$ gives
\begin{equation}
  \sum_{i=1}^N\omega_i = \langle \mu,\nu\rangle.
\end{equation}
It was shown in \cite{VW:} that the effect of the superpotential is to
contribute $-2\sum\omega_i$ to $d$. Thus we have
\begin{equation}
 d=n-2\langle \mu,\nu\rangle,		\label{eq:d}
\end{equation}
in agreement with the conjecture in \cite{BB:mir}.\footnote{Except
that there appears to be a typographical error in conjecture (2.17) of
\cite{BB:mir}.}

For the cases considered in \cite{AGM:II} based upon the construction
of \cite{Bat:m} we had $\langle\mu,\nu\rangle=1$. This then is a
generalization. It should be noted that this more generalized picture
could have been deduced directly by applying the toric language to Witten's
formulation of \cite{W:phase} although historically it was first
written in the form of \cite{Boris:m} where it was used specifically
for conjecturing the mirror map for complete intersections in toric
varieties.

To summarize so far, all the data we require to build an abelian
gauged linear \sm\ of the form studied in \cite{W:phase} is the matrix
$\alpha_{ij}$. To provide a consistent model for a conformal field
theory we demand that this matrix be compatible with $\mu$ and $\nu$
and be consistent with
the existence of ${\bf N}$ in the form of (\ref{eq:Acvx}). Once we
have this information we may apply
the technology of \cite{W:phase,AGM:II} to determine the geometry of
the various phases in the moduli space of K\"ahler forms. This is most
easily determined in terms of triangulations of the point set $\cA$.

There is one more piece of information we will need before moving on
to some examples concerning orbifolding. The toric variety $\Vbig$ is
acted upon by $(\C^*)^n$. It is simple in toric geometry to describe
the orbifold of $\Vbig$ by a discrete subgroup of this $(\C^*)^n$.
Consider the affine coordinates introduced by (\ref{eq:aff}). Let us
consider the element, $g\in(\C^*)^n$ which acts by
\begin{equation}
  g:(\zeta_1,\zeta_2,\ldots,\zeta_n)\mapsto(e^{2\pi ig_1}\zeta_1,
    e^{2\pi ig_2}\zeta_2,\ldots,e^{2\pi ig_n}\zeta_n) \label{eq:iden}
\end{equation}
where $0\leq g_j<1$. We can see (for more details consult  \cite{Reid:yp})
that dividing $\Vbig$ by the group generated by $g$ is equivalent to
replacing the lattice ${\bf N}$ by a lattice generated by ${\bf
N}$ and the vector $(g_1,g_2,\ldots,g_n)\in{\bf N}_\R$. The reason for this is
that lattice points $p$ in ${\bf N}$ represent one (complex) parameter group
actions on the toric variety
\begin{equation}
  p:(\zeta_1,\zeta_2,\ldots,\zeta_n)\mapsto(\lambda^{p_1}\zeta_1,
    \lambda^{p_2}\zeta_2,\ldots,\lambda^{p_n}\zeta_n).
\end{equation}
For points $p$ whose components are non-integral, such a map is
only well defined if certain global identifications are made on
the $(\zeta_1,\zeta_2,\ldots,\zeta_n)$. In particular, one directly sees
that taking $p$ to be $(g_1,g_2,\ldots,g_n)$ requires the desired
identification of (\ref{eq:iden}).

\section{Applications}   \label{s:appl}

Let us now illustrate the general method of the previous section by
applying it to various examples. The possibilities offered by this
formulation appear to be very rich but we select here a few key
examples to emphasize points relevant to our discussion.

\subsection{The Hypersurface Case}   \label{ss:hyp}

Suppose that $\langle \mu,\nu\rangle=1$. In this case it is easy to
see that $\nu\in\cA$ and that this point lies properly in the interior
of the convex hull of $\cA$ (since $\langle\beta_v,\nu\rangle$ is
strictly positive). One possible triangulation of the point set $\cA$
thus consists of drawing lines from $\nu$ to each point on the vertices of
the convex hull and filling this skeleton in with a suitable set of
simplices to form a triangulation. The resultant set forms a complete fan of
dimension $n-1$ with center $\nu$. This fan $\delta$ corresponds to a
compact $(n-1)$-dimensional toric sub-variety $\Vlit$ of
$\Vbig$. Let us denote by $p$ the homogeneous coordinate
corresponding to $\nu$. We see from (\ref{eq:mon}) that every term
in the superpotential appears linearly in $p$. Thus we may write
$W=pG$ where $G$ is a function of the $N-1$ homogeneous coordinates
describing $\Vlit$. Thus the condition $\partial W/\partial p=0$
implies $G=0$ --- i.e., we are on a hypersurface within $\Vlit$. For a
generic $G$, the other derivatives of $W$ imply that $p=0$.

The target space, $X$, is now a hypersurface within $\Vlit$ which itself
has dimension $n-1$. $X$ is thus of dimension $n-2$. The equation
(\ref{eq:d}) tells us that $d=n-2$.
In fact $X$ is an anticanonical divisor of $\Vlit$ \cite{Bat:m} and
is thus a \CY\ space of $d$ dimensions.
Note that $X$ may
not be smooth but these singularities can often be removed by further
refinements of the fan $\delta$. Actually, in the case $d\leq3$ the
singularities may always be removed in this way.\footnote{This is
because Gorenstein singularities can only be terminal in more than 3
dimensions \cite{Reid:mm}.}

For the case $\langle \mu,\nu\rangle=1$ we therefore always have a
``\CY'' phase. That is, some limit in the moduli space where we may go
to build some non-linear \sm\ of the conformal field theory (although
in case $d>3$ we may have to include considerations such as terminal
orbifold singularities in our model). The case considered here is
basically of the type studied in \cite{Bat:m,AGM:II} as shown in
\cite{BB:mir}. It also includes the example of the model with the \LG\
phase in $\C^n/\Z_n$ and the \CY\ hypersurface in $\P^{n-1}$ discussed above.

\subsection{The Mirror of the Z-orbifold}
		\label{ss:Z}

We now turn to the issue of rigid Calabi-Yau spaces and their mirrors.
For concreteness we focus on the Z-orbifold of \cite{CHSW:}. Recall
that this is the torus of six real dimensions divided by a diagonal
$\Z_3$ action.
It has 36 (1,1)-forms (9 from the original torus and 27 associated
with blow up modes) and no (2,1)-forms. It is therefore rigid.
Using the construction of \cite{GP:orb}, it was shown in
\cite{AL:geom} how to construct the Z-orbifold
in terms of an orbifold of a Gepner model \cite{Gep:}. To
phrase this
more carefully allowing for the phase structure, one
builds a conformal field theory as an orbifold of a Gepner model which
may be deformed via marginal operators to a theory corresponding to a
\sm\ whose target space is the blown-up Z-orbifold. It was also shown
how to build a conformal field theory giving the mirror of the above
theory, also as a orbifold of the Gepner model.

The Gepner model itself is believed to be equivalent to an orbifold of
a \LG\ theory. In the case under consideration (the ${\bf 1}^9$ model)
the configuration space of this \LG\ orbifold theory is $\C^9/\Z_3$. The
space $\C^9$ is a
toric variety with $n=9$ described simply by the fan consisting of one
cone, $\sigma$, isomorphic to the positive quadrant of $\R^9$. That is, $\cA$
consists of the points
$(1,0,0,0,0,0,0,0,0),(0,1,0,0,0,0,0,0,0),
\ldots,(0,0,0,0,0,0,0,0,1)$. The required $\Z_3$ quotient is performed
by adding the generator
\begin{equation}
  g_1=(\ff13,\ff13,\ff13,\ff13,\ff13,\ff13,\ff13,\ff13,\ff13),
		\label{eq:g1Z}
\end{equation}
to the integral lattice of $\R^9$. It was shown in \cite{AL:geom} that
the mirror of the Z-orbifold was obtained by dividing by a further
$\Z_3$ (i.e., taking a $\Z_3$ orbifold of the Gepner model) given by the vector
\begin{equation}
  g_2=(0,0,0,\ff13,\ff13,\ff13,\ff23,\ff23,\ff23),
\end{equation}
to give the required ${\bf N}$-lattice.
We may apply a $Gl(9,\R)$ transformation to ${\bf N}_\R$ to rotate
${\bf N}$ back into the standard integral lattice. This will act on
$\sigma$ so that it is no longer the positive quadrant. One choice of
transformation leaves $\sigma$ generated by
\begin{equation}
  \eqalignsq{
   \alpha_1&=(3,0,0,1,1,1,-1,-1,-3)\cr
   \alpha_2&=(0,1,0,0,0,0,0,0,0)\cr
   \alpha_3&=(0,0,1,0,0,0,0,0,0)\cr
   \alpha_4&=(0,0,0,1,0,0,0,0,0)\cr
   \alpha_5&=(0,0,0,0,1,0,0,0,0)\cr
   \alpha_6&=(0,0,0,0,0,1,0,0,0)\cr
   \alpha_7&=(0,0,0,0,0,0,1,0,0)\cr
   \alpha_8&=(0,0,0,0,0,0,0,1,0)\cr
   \alpha_9&=(0,-1,-1,-2,-2,-2,0,0,3).\cr}	\label{eq:AZ}
\end{equation}
These 9 points lie in the hyperplane defined by
$\mu=(3,1,1,1,1,1,1,1,3)$. It is a simple matter to show that the dual
cone gives $\nu=(1,0,0,0,0,0,0,0,0)$ so that $\cA$ has the required properties.

The important property of this model stems from the fact that the
points $\alpha_1,\ldots,\alpha_9$ form the vertices of a simplex with
{\em no\/} interior points lying on the lattice ${\bf N}$. That is,
the set $\cA$ consists only of those points listed in (\ref{eq:AZ}).
Thus the only triangulation of $\cA$ consists of this simplex! This
model has $\Vbig\cong\C^9/(\Z_3\times\Z_3)$ with superpotential
\begin{equation}
  W = a_1x_1^3+a_2x_2^3+\ldots+a_9x_9^3+a_{10}x_1x_2x_3+\ldots.	\label{eq:WZ}
\end{equation}
The critical point of $W$ is the origin. Thus we have an orbifold of a
\LG\ theory as expected. Since there is no other triangulation of
$\cA$ there is no other phase and, in particular, {\it  no \CY\ phase}.
Since $\langle \mu,\nu\rangle=3$ we are not in conflict with the
section \ref{ss:hyp}. As expected we see that $d=3$ in agreement with the
fact that this theory is the mirror of a smooth \CY\ threefold (i.e.,
the blow-up of the Z-orbifold).

Thus, by properly understanding the full content of mirror symmetry ---
as a symmetry between the moduli spaces of $N = 2$ superconformal
theories ---
we see that there is no puzzle regarding the mirror of a rigid Calabi-Yau
manifold. The mirror description simply does not have a Calabi-Yau phase
and hence the absence of a K\"ahler form causes no conflict.

It is important to realize that we have all the information we need to
study this model without recourse to finding some other effective
target space geometry. In particular, deformations of complex
structure are achieved by deforming the $a_v$ parameters in
(\ref{eq:WZ}) in the usual way and one may then use mirror symmetry to
study the moduli space of K\"ahler forms of the Z-orbifold as was done
in \cite{Drk:Z}.\footnote{Note that the periods deduced in
\cite{Drk:Z} can be determined from the analysis of the Picard-Fuchs
equation as we mention briefly later.}

The lack of a \CY\ phase appears due to the existence of {\em terminal\/}
singularities in algebraic geometry as we now discuss (see also
\cite{Reid:yp} for a more thorough account).
In section \ref{s:ph} we discussed the case
of a Landau-Ginzburg theory in $\C^n/\Z_n$. In this case, the $\Z_n$
symmetry is generated by $(\ff1n,\ff1n,\ldots)$. This singularity may
be blown-up to give the canonical line bundle over $\P^{n-1}$. This
smooth space has trivial canonical class. Thus the singularity
$\C^n/\Z_n$ may be blown up without adding something non-trivial into
the canonical class. Such a blow-up mode is always visible in the
associated conformal field theory as a truly marginal operator since
it may be regarded as a deformation of the K\"ahler form.

A terminal singularity is a singularity which cannot be resolved (or
even partially resolved) without adding something non-trivial to the
canonical class. The singularity $\C^9/\Z_3$ generated by $g_1$ of
(\ref{eq:g1Z}) is precisely such a singularity. As such, from the
conformal field theory point of view, it is ``stuck''. This agrees
with the fact that the Gepner model contains no marginal operators
corresponding to deformations of the K\"ahler form.

One can go ahead and blow-up the $\C^9/\Z_3$ singularity if one really
desires some smooth manifold. There is no unique prescription for this
but one may, for example, form the space $\O_{\P^8}(-3)$. This is a
line bundle over $\P^8$ with $K<0$ (i.e., $c_1>0$). The homogeneous
coordinates of
this projective space may be given by the coordinates of the original
$\C^9$. The superpotential of the Landau-Ginzburg theory is cubic in
these fields and so one might try to associate this model to the cubic
hypersurface in $\P^8$. This is the essence of the construction of
\cite{Drk:Z,Schg:gen}. Note that in the language of this paper,
we no longer satisfy (\ref{eq:Q0}) and so our field theory is expected
to have
undesirable properties in the infrared limit.

When we try to describe the mirror of the Z-orbifold, the situation
becomes even worse. The second $\Z_3$ quotient given by $g_2$ induces
further terminal quotient singularities on $\P^8$ which require
considerably more to be added to the canonical class. We hope the
reader sees that this procedure of forcing a smooth geometrical
interpretation when terminal singularities appear is completely
unnatural when written in terms of the underlying conformal field
theory and it is unnecessary when one adopts the phase picture. The
mirror of the Z-orbifold need only be described as an orbifold of a
Landau-Ginzburg theory in $\C^9$.

We should add that the construction of \cite{Set:sup} should be
expected to overcome the renormalization group flow problem inherent
in the above
hypersurface in $\P^8$ of \cite{Drk:Z,Schg:gen}. In the construction
of \cite{Set:sup} one adds
ghost fields to reduce the effective dimension of the target space
back down to that of $d$. Assuming this is the case, this target space
with ghosts can be proposed as a good geometric interpretation of the
conformal field theory. It should be pointed out however that such
geometric interpretations are probably highly ambiguous. That is, one
conformal field theory can be given many interpretations. This occurs
in \cite{Set:sup} where constructions of K3 conformal field theories
are given in
terms of a 4 complex dimensional space with ghosts whereas the
complete moduli
space is already understood completely in terms of K3 surfaces
\cite{AM:K3p}. In fact, it is probable that any geometric model may be
blown-up to give $K<0$ and then nonzero contributions the the
$\beta$-function be cancelled by adding suitable extra fields. Since
there are an infinite number of such blow-ups for any model there is
the possibility of ascribing an infinity of geometric interpretations
of this form.

Finally note that it might be possible to associate some geometry with
the case discussed in this section by considering orbifolds with
discrete torsion \cite{Berg:dt}. Since we do not understand precisely
how to relate quotient singularities with discrete torsion to
classical singularities we will not discuss this interpretation here.

\subsection{A Case with $h^{1,1}=1$}
			\label{ss:h1}

The above example may be considered rather trivial in that our phase
space was zero dimensional, i.e., consisted of only one point. Let us
now give a less trivial example which still has no \CY\ phase.

Consider dividing $\C^9$
by the group $\Z_4\times\Z_4$ groups generated by
\begin{equation}
\eqalign{g_1&=(\ff14,\ff14,\ff12,\ff14,\ff14,\ff12,\ff14,\ff14,\ff12)\cr
  g_2&=(\ff14,\ff34,0,0,0,0,0,0,0).\cr}
		\label{eq:gZ4}
\end{equation}
Using the arguments of \cite{GP:orb,AL:geom} one may show that the
\LG\ theory in this space is
the mirror of the orbifold $T^3/(\Z_4\times\Z_2)$ where $T$ is a
complex torus, the $\Z_4$ group is generated by
$(z_1,z_2,z_3)\mapsto(iz_1,-iz_2,z_3)$ and $\Z_2$ by
$(z_1,z_2,z_3)\mapsto(z_1,-z_2,-z_3)$, where $z_i$ are the complex
coordinates on the tori. The $(2,1)$-form $dz_1\wedge dz_2\wedge d\bar
z_3$ is
invariant under this group. Indeed $h^{2,1}$ for this orbifold is equal
to 1. Thus we expect the case in question to have $h^{1,1}=1$.

The point set $\cA$ corresponding to such a space is given by $n=9$ and
\begin{equation}
  \eqalignsq{
   \alpha_1&=(4,-3,0,0,0,0,0,0,0)\cr
   \alpha_2&=(0,1,0,0,0,0,0,0,0)\cr
   \alpha_3&=(0,0,1,0,0,0,0,0,0)\cr
   \alpha_4&=(0,0,0,1,0,0,0,0,0)\cr
   \alpha_5&=(0,0,0,0,1,0,0,0,0)\cr
   \alpha_6&=(0,0,0,0,0,1,0,0,0)\cr
   \alpha_7&=(0,0,0,0,0,0,1,0,0)\cr
   \alpha_8&=(-4,2,-2,-1,-1,-2,-1,4,-2)\cr
   \alpha_9&=(0,0,0,0,0,0,0,0,1)\cr
   \alpha_{10}&=(3,-2,0,0,0,0,0,0,0)\cr
   \alpha_{11}&=(2,-1,0,0,0,0,0,0,0)\cr
   \alpha_{12}&=(1,0,0,0,0,0,0,0,0),\cr}	\label{eq:Ao}
\end{equation}
with $\mu=(1,1,1,1,1,1,1,3,1)$ and $\nu=(0,0,0,0,0,0,0,1,0)$.
Therefore this theory has $d=3$ again.

The points $\alpha_1,\ldots,\alpha_9$ form a simplex with
$\alpha_{10},\alpha_{11},\alpha_{12}$ positioned along the edge
joining $\alpha_1$ and $\alpha_2$. Thus all the interesting part of
the point set as regards triangulations is contained in this line
$\alpha_1\alpha_2$:
\begin{equation}
\setlength{\unitlength}{0.007in}%
\begin{picture}(330,30)(155,635)
\thinlines
\put(400,660){\circle{10}}
\put(160,660){\circle*{10}}
\put(480,660){\circle*{10}}
\put(320,660){\circle{10}}
\put(240,660){\circle{10}}
\put(160,660){\line( 1, 0){320}}
\put(315,635){\makebox(0,0)[lb]{$\alpha_{11}$}}
\put(155,635){\makebox(0,0)[lb]{$\alpha_1$}}
\put(235,635){\makebox(0,0)[lb]{$\alpha_{10}$}}
\put(395,635){\makebox(0,0)[lb]{$\alpha_{12}$}}
\put(475,635){\makebox(0,0)[lb]{$\alpha_2$}}
\end{picture}
\end{equation}
The points $\alpha_{10},\alpha_{11},\alpha_{12}$ may, or may not be
included in the triangulation (and are hence shown as circles rather
than dots).

If none of the points $\alpha_{10},\alpha_{11},\alpha_{12}$ are
included in the triangulation, we have one simplex with vertices
$\alpha_1,\ldots,\alpha_9$ and the associated toric variety is
$\C^9/(\Z_4\times\Z_4)$ as expected. If all these points are included
in the triangulation we have 4 simplices. The resulting space is a
partial resolution of the $\C^9/(\Z_4\times\Z_4)$ space. The
exceptional divisor introduced is a ``plumb product'' of three $\P^1$
spaces. Each of the points $\alpha_{10},\alpha_{11},\alpha_{12}$ may be
taken to correspond to one of these $\P^1$ components. This is shown in
figure \ref{fig:Z4b}. The black dot on the left hand side shows the
isolated singularity. On the right hand side the singularity (which is
now terminal) covers the whole exceptional divisor.
Clearly, other triangulations represent intermediate steps in this blow-up.
\iffigs
\begin{figure}
  \centerline{\epsfxsize=12cm\epsfbox{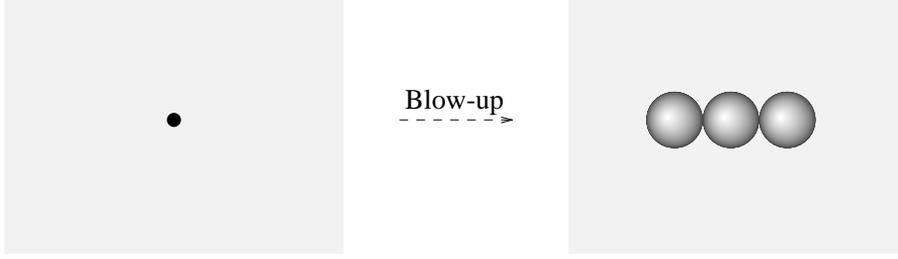}}
  \caption{Maximal partial resolution of $\C^9/(\Z_4\times\Z_4)$ with $K=0$.}
  \label{fig:Z4b}
\end{figure}
\fi

Let us now analyze the critical point set of $W$. Finding $\cB$ we
determine from (\ref{eq:mon}) that
\begin{equation}
W = a_1x_1^4x_{10}^3x_{11}^2x_{12}+a_2x_2^4x_{10}x_{11}^2x_{12}^3+
    a_3x_3^2+a_4x_4^4+a_5x_5^4+a_6x_6^2+a_7x_7^4+a_8x_8^4+a_9x_9^2+
    \ldots		\label{eq:WZ4}
\end{equation}
In total there are 87 points in $\cB$ but we need only consider the
above terms with nonzero $a_1,\ldots,a_9$ for a sufficiently generic
$W$.

Consider the maximal triangulation. This includes all three points
$\alpha_{10},\alpha_{11},\alpha_{12}$. Since $N=12$ we need to remove
the set $F_\Delta$ given by (\ref{eq:Fset}) from $\C_{12}$. This
amounts in imposing $x_2x_{11}x_{12}\neq0$ or $x_1x_2x_{12}\neq0$ or
$x_1x_2x_{10}\neq0$ or $x_1x_{10}x_{11}\neq0$. We also wish to impose
$\partial W/\partial x_i=0$ for $i=1,\ldots,12$. It is
straight-forward to show that these conditions require
\begin{equation}
  \eqalign{x_3=x_4=x_5=x_6=x_7&=x_8=x_9=x_{11}=0\cr
    x_1&\neq0\cr
    x_2&\neq0\cr}
\end{equation}
and that $x_{10}$ and $x_{12}$ cannot both be zero simultaneously. As
$N-n=3$ we have three $\C^*$ actions to divide this subspace of $\C^{12}$
by. Two may be used to fix $x_1$ and $x_2$ to specific values. The
other $\C^*$ may be used to turn $x_{10}$ and $x_{12}$ into
homogeneous coordinates parametrizing $\P^1$. The vacuum is thus
$\P^1$. One may also determine the superpotential in this vacuum to
show that we have a \LG\ theory fibered over this $\P^1$ to
obtain the
familiar hybrid-type models of \cite{W:phase,AGM:II}. One may also
show that the fiber has a $\Z_4$-quotient singularity at the zero section.

In terms of the ambient toric variety $V_\Delta$, what we have just
described in the previous paragraph is the $\P^1$ that appears in the
middle of the chain of three
$\P^1$'s on the right in figure \ref{fig:Z4b}. Thus although
$\Vbig$ appears to have three degrees of freedom for the K\"ahler
form --- giving the three independent sizes of the three $\P^1$'s,
only one makes it down to $X$, the critical point set of $W$.
Therefore $X$ only has one K\"ahler-type deformation. Sometimes
additional modes appear in the fibre for these hybrid models but in
this case the \LG\
fibre contains no twist fields with the correct charges to be considered
a (1,1)-form.
We will therefore assert that $h^{1,1}(X)=1$. Thus we are in
agreement with the assertions concerning the mirror space at the start
of this section.

Analyzing the other possible triangulations we find that we reproduce
one of the two phases we know about --- either the
\LG\ orbifold in $\C^9/(\Z_4\times\Z_4)$ or the hybrid model over
$\P^1$. The points $\alpha_{10}$ and $\alpha_{12}$ may be ignored when
considering $X$. Thus we have constructed a model with a non-trivial
phase diagram --- there are 2 phases --- but neither is a \CY\ space.

In general there is a homomorphism:
\begin{equation}
  \kappa:H^{1,1}(V_\Delta)\to H^{1,1}(X).
		\label{eq:kappa}
\end{equation}
In general however $\kappa$ is neither injective nor surjective. The
example in this section shows a failure of injectivity since
$h^{1,1}(V_\Delta)=3$ and $h^{1,1}(X)=1$.
In the more simple case of
$\langle\mu,\nu\rangle=1$ it was shown in \cite{AGM:mdmm} that the
kernel of $\kappa$ was described by points in the interior of
co-dimension one faces of the convex hull of $\cA$. In the case
described in this section we see that such a simple criterion cannot
be used --- all the points $\alpha_{10},\alpha_{11},\alpha_{12}$ lie
in a co-dimension 7 face and yet $\alpha_{10}$ and $\alpha_{12}$
contribute to the kernel and $\alpha_{11}$ survives through to
$H^{1,1}(X)$. At this point in we know of no simple method of
determining the image of $\kappa$ except to explicitly calculate the
critical point set of $W$ on a case by case basis.

Let us conclude this section by discussing the short-comings of
analyzing this model in terms of the ``generalized \CY\ manifolds'' of
\cite{Schg:gen}. The $\Z_4$ singularity in $\C^9$ generated by $g_1$
of (\ref{eq:gZ4})
may be partially resolved by a line bundle over the weighted projective space
$\P^8_{\{1,1,2,1,1,2,1,1,2\}}$. The resultant space has $K<0$.
The ``generalized \CY\ manifold'', $R$, would be identified as the
hypersurface in this weighted projective space given by the vanishing
of (\ref{eq:WZ4}) with $x_1,\ldots,x_9$ taken to be the
quasi-homogeneous coordinates and $x_{10}=x_{11}=x_{12}=1$.
The
$\Z_4$-action of $g_2$ acts on $R$ to
induce $\Z_4$-quotient singularities over some subspace of codimension
two. These latter
singularities are not terminal and may be resolved without adding
anything further to $K$. In fact, resolving these latter singularities
may be achieved by introducing the points
$\alpha_{10},\alpha_{11},\alpha_{12}$ into the toric fan.

It is easy to see that something similar will happen in general. That
is, any $K=0$ toric resolutions we may perform in $\Vbig$ may also be
performed after blowing up any terminal singularities in $\Vbig$. It follows
that the points in the interior of the convex hull of $\cA$ may
be counted by analyzing singularities which may be locally resolved
with $K=0$ in $R$. (Note that this is a
rather inefficient way of proceeding in our picture --- one may as
well just analyze the singularities in $\Vbig$ without any destroying
the $K=0$ condition.) This observation sheds light on a conjecture in
\cite{Schg:fno} that $h^{1,1}(X)$ could be determined by counting the
contribution to $h^{1,1}$ of any resolutions of singularities within
the $R$. We see now that this will count
$h^{1,1}(V_\Delta)$ which is, in general, not equal to $h^{1,1}(X)$.
Thus this conjecture is false. In the example above, counting this way
would imply that $h^{1,1}(X)=3$.

With regards to determining $h^{1,1}(X)$, it appears hard to save the
construction of \cite{Set:sup} from a
similar fate. The problem is that the divisors associated with
$\alpha_{10}$, $\alpha_{11}$ and $\alpha_{12}$ appear on equal footing
in $R$. Thus unless some unsymmetric rules are devised for resolving
canonical singularities in superspace one cannot obtain
the correct answer $h^{1,1}(X)=1$.

\subsection{A Case with $X=\hbox{\bigbbbfont P}^3$}
		\label{ss:P3}

One might be led to suspect the following to be the general picture
for the geometric interpretation of an $N$=2 superconformal field
theory. Either $X$ is a \CY\ space and the string is free to move
within $X$ and there are no massless modes normal to $X$, or $X$ is a
space of dimension $<d$ in which the string is free to move and there
are massless modes governed by some superpotential
normal to $X$ inside some bigger ambient space
$\Vbig$ containing $X$. We now show an example (which also appeared in
\cite{W:phase}) which is an exception to this.

Consider the following point set for $\cA\,$:
\begin{equation}
  \eqalignsq{
   \alpha_1&=(1,0,0,0,0,0,0,1,0,0,0)\cr
   \alpha_2&=(0,1,0,0,0,0,0,1,0,0,0)\cr
   \alpha_3&=(0,0,1,0,0,0,0,0,1,0,0)\cr
   \alpha_4&=(0,0,0,1,0,0,0,0,1,0,0)\cr
   \alpha_5&=(0,0,0,0,1,0,0,0,0,1,0)\cr
   \alpha_6&=(0,0,0,0,0,1,0,0,0,1,0)\cr
   \alpha_7&=(0,0,0,0,0,0,1,0,0,0,1)\cr
   \alpha_8&=(-1,-1,-1,-1,-1,-1,-1,0,0,0,1)\cr
   \alpha_9&=(0,0,0,0,0,0,0,1,0,0,0)\cr
   \alpha_{10}&=(0,0,0,0,0,0,0,0,1,0,0)\cr
   \alpha_{11}&=(0,0,0,0,0,0,0,0,0,1,0)\cr
   \alpha_{12}&=(0,0,0,0,0,0,0,0,0,0,1).\cr}
		\label{eq:AP3}
\end{equation}
Thus $\mu=(0,0,0,0,0,0,0,1,1,1,1)$. One can also determine $\cB$ with
a little effort and
find $\nu=(0,0,0,0,0,0,0,1,1,1,1)$. Thus $d=3$ again. The
superpotential $W$ may be written
\begin{equation}
  W = x_9G_1 + x_{10}G_2 + x_{11}G_3 + x_{12}G_4,
\end{equation}
where the $G_k$ are generic homogeneous polynomials of total degree two in
$x_1,\ldots,x_8$.

There are two triangulations of the point set $\cA$. The first
consists of taking 8 simplices each of which has
$\alpha_9,\ldots,\alpha_{12}$ as 4 of its vertices with the other 7
vertices taken from the set $\{\alpha_1,\ldots,\alpha_8\}$. In terms
of $F_\Delta$ this amounts to removing the point $x_1=\ldots=x_8=0$
from consideration. Restricting to the critical point set of $W$ forces
$x_9=\ldots=x_{12}=0$ and $G_1=\ldots=G_4=0$. Dividing out by the
single required $\C^*$-action forms $\P^7$ with homogeneous
coordinates $x_1,\ldots,x_8$. Thus $X$ is the intersection of 4
quadric equations $G_k=0$ in $\P^7$. This is a known \CY\ space dating
back to \cite{CHSW:}.

The other triangulation consists of 4 simplices with 8 vertices given
by $\alpha_1,\ldots,\alpha_8$ with the other 3 taken from the set
$\{\alpha_9,\ldots,\alpha_{12}\}$. This amounts to removing
$x_9=\ldots=x_{12}=0$ from consideration. Restricting to the critical
point set of $W$ forces $x_1=\ldots=x_8=0$. Now the $\C^*$-action may
be used to form $\P^3$ with homogeneous coordinates
$x_9,\ldots,x_{12}$. Thus $X$ in this phase is $\P^3$.

Our phase diagram consists of two phases --- both of which have the
dimension of $X$ equal to $d$. One phase is a \CY\ manifold with
$h^{1,1}=1$ and $h^{2,1}=65$ which we
understand. The other phase is $\P^3$. The reader might be alarmed at
the appearance of the latter since $\P^3$ is not a \CY\ space and
lacks a nonvanishing holomorphic 3-form for example.
The resolution is as follows. Although we have
correctly identified the vacuum of the field theory as $X$ we have to
be a little careful in declaring it to be the effective target space
of a conformal field theory. Let us consider the variables
$\alpha_1,\ldots,\alpha_8$ which we forced to zero. The superpotential
is quadratic in these variables so we certainly haven't missed any
massless degrees of freedom (which would only add to our troubles by
increasing $d$ anyway). The point is that there is actually a
$\Z_2$-quotient singularity coming from the identification of
homogeneous coordinates in $\Vbig$ to affine ones. Thus we have a
fibration of a \LG\ orbifold theory over $\P^3$ which may appear
trivial in that the superpotential is quadratic but we may expect
twist fields is add to our spectrum. In particular we expect to have an
analog to a Calabi-Yau $H^{3,0}$ mode (i.e. a field of charge (3,0) under
the $U(1)_L \times U(1)_R$ of the superconformal algebra) coming
from such twisted sectors. Of course, such a mode cannot be given
a literal geometric interpretation in terms of a (3,0)-form.

It is interesting also to ask how literally we can take this $\P^3$ to
be a target space for the conformal field theory. To find the actual
size of truly conformally invariant \sm\ target space one needs to solve the
Picard-Fuchs system as described in \cite{AGM:sd}. We will not
present the details here since they are rather lengthy but we may
quickly summarize as follows. One solves equations (42) of
\cite{AGM:sd} where the ``$\beta$'' vector of this system is set equal to
$-\nu$. (One could then count rational curves on this \CY\ space if
one so desired.) The complexified K\"ahler form $B+iJ$ of the \CY\
phase can then be analytically continued into $\P^3$ phase (which is
most easily done by the method of \cite{me:min-d}). Taking $z$ to be
the local coordinate on the moduli space where $z=0$ corresponds to
the limit point in the $\P^3$ phase we obtain
\begin{equation}
  B+iJ=-\ff12 -\frac{3\pi i}{2\log(z)} +O(\log(z)^{-2}).
\end{equation}
Thus $J\geq 0$ in the region near the limit point $|z|\ll1$. The
effective size of target space is very small as
$|z|\to0$. In other words, the effect of integrating out the massive
modes in the linear \sm\ of \cite{W:phase} has caused an infinite
renormalization of the ``$r$'' parameter (unlike what is believed to
happen for the \CY\ phase).

To summarize we see that the phase picture can produce phases with
dimension equal to $d$ which do not correspond to \CY\ non-linear \sm
s. To understand these phases more completely will require a better
understanding of the hybrid models.


\section{Conclusions}    \label{s:conc}

Geometrical methods have proven themselves to be a powerful conceptual
and calculational tool in understanding the physical content of
certain conformal theories and their associated string models. As
such, it is a worthwhile task to gain as complete an understanding as
possible of the geometrical status of conformal field theories,
especially for the case of $N = 2$ worldsheet supersymmetry relevant
for spacetime supersymmetric string models. The phase structure of
such $N = 2$ models, as found in
\cite{W:phase,AGM:II}, goes a long way towards capturing the full
geometric content of these theories, and, in particular, certainly
provides the correct framework for discussing mirror symmetry. In this
paper we have used this phase structure analysis to address certain
previously puzzling issues regarding the geometrical content of
certain theories. In particular, at first sight mirror symmetry seems
to come upon the puzzle regarding the identity of the mirror of a
rigid manifold.  We have seen, though, that this appears to be a
puzzle only because the question itself is not phrased in the correct
context. That is, mirror symmetry tells us that certain {\it a
priori\/} distinct pairs of families of conformal theories actually
are composed of isomorphic members. When the phase structure of each
family in such a pair contains a Calabi-Yau \sm\ region, then these
Calabi-Yau's form a mirror pair. However, in certain cases, at least
one of the families does {\it not\/} have a Calabi-Yau \sm\ region. In
such cases mirror symmetry will simply not yield a mirror pair of
Calabi-Yau manifolds. A family which has a rigid Calabi-Yau phase, as
we have seen, provides one such example --- the mirror family does not
have a Calabi-Yau region. Thus the absence of a K\"ahler form for the
mirror is not an issue. The mirror moduli space has no Calabi-Yau
phase and hence does not require a K\"ahler form.  The previous puzzle
disappears, therefore, when the question is phrased in the correct
context.

Beyond the rigid case, we have also seen that even when there is a
conformal mode that can play the part of a K\"ahler form, there need
not be a Calabi-Yau phase on which it can realize this potential. So,
whereas in the previous problem we resolved the issue of a
``Calabi-Yau in search of a K\"ahler form'' here we have ``a K\"ahler
form in search of a Calabi-Yau''.  We have established that there are
examples in which there simply is no Calabi-Yau to be found.

It is worth noting that the map $\kappa$ in (\ref{eq:kappa}) is
neither injective or surjective. We saw the effect on this map not
being injective in section \ref{ss:h1}. The failure of surjectivity
shows that some (1,1)-forms on $X$ do not come from the toric ambient
space $\Vbig$. In the case of models of the form discussed in section
\ref{ss:hyp} it is still possible to count the number $h^{1,1}(X)$
because of the properties of hypersurfaces \cite{Bat:m,AGM:mdmm}. In
the cases
considered here however we deal with more general complete
intersections. The problem of counting $h^{1,1}(X)$ in this context
is the mirror of the problem of counting $h^{2,1}$ for
the mirror model. The analogue of the fact that $\kappa$ is not
an isomorphism is the fact that deforming the polynomial giving the
superpotential is not the same as deformation the complex structure.
It would be interesting to see if methods
along the lines of \cite{GH:poly} could be applied in this context to
determine the Hodge numbers of $X$.

An interesting question, to which we do not know the answer, is whether
there are examples in which {\it neither\/} family in a mirror pair has
a Calabi-Yau region. Such an example would establish that there are
$N = 2, c = 3d$ conformal theories which are not interpretable in terms
of Calabi-Yau compactifications (or analytic continuations thereof).
To answer this question is difficult because of the failure
of the surjectivity of $\kappa$. One may find mirror pairs of
orbifolds of the Gepner model, ${\bf 1}^9$, for which neither has an obvious
\CY\ interpretation. An example with $h^{1,1}$ and $h^{2,1}$ equal to
4 and 40 was mentioned in \cite{AL:geom}. This is a good candidate for
a situation where neither of the mirror partners have a \CY\ phase
(despite the assertions of \cite{AL:geom}). Unfortunately if $X$ is
the model with $h^{1,1}=4$ then the image of $\kappa$ is trivial,
i.e., none of the (1,1)-forms come from $\Vbig$. Because of this the
methods of this paper cannot be used to draw any conclusions regarding
the lack of \CY\ phase for this example.


\section*{Acknowledgements}

It is a pleasure to thank D. Morrison and R. Plesser
for useful conversations.
The work of P.S.A. is supported by a grant from the National
Science Foundation. The work of B.R.G. is supported by
a National Young Investigator Award, the Alfred P. Sloan Foundation and
the National Science Foundation.


\section*{Note Added}

The mirror of the example of section \ref{ss:P3} was studied in
\cite{lots:per} where it was discovered that there is an extra $\Z_2$
symmetry in the moduli space. This should act on the moduli space in
section \ref{ss:P3} to identify the two phases with each other. This
may be viewed as a new kind of $R\leftrightarrow1/R$ symmetry.


\end{document}